\documentclass[preprint,number,sort&compress]{elsarticle}
\usepackage{graphicx} \usepackage{subfigure} \usepackage{amssymb}
\usepackage{verbatim} \usepackage{amsmath} \usepackage{amscd}
\usepackage{amssymb} 
\usepackage{epsfig}     

\newcommand{\rd}[1]{\mathop{\mathrm{d}#1}}

\newcommand{\met}{g_{\mu\nu}}

\newcommand{\p}{x_+}
\newcommand{\m}{x_-}
\newcommand{\g}{g_{+-}}
\newcommand{\diff}{\mathcal{L}_{\xi}}
\newcommand{\exo}{\theta_{(l)}}
\newcommand{\ar}{a_{\Delta}}	
\newcommand{\AdS}{\mathrm{AdS}_3}

\begin{document}

\title{Hawking Radiation and Entropy from Horizon Degrees of Freedom}

\author[hc]{Hyeyoun Chung}  
\ead{hyeyoun@physics.harvard.edu.}
\address[hc]{Jefferson Physical Laboratory, Harvard University,\\ 17 Oxford St., Cambridge, MA 02138, USA}

\date{\today} \begin{abstract}We study the thermodynamic properties of horizons using the dynamical description of the gravitational degrees of freedom at a horizon found in a previous work. We use the action of the horizon degrees of freedom to calculate the horizon entropy using the Cardy formula, and obtain the expected Bekenstein-Hawking entropy. We also couple the gravitational degrees of freedom at the horizon to a classical background scalar field, and show that Hawking radiation is produced.\end{abstract}



\maketitle 

\section{Introduction}

Ever since the work of Bekenstein and Hawking in the 1970s, which established that the laws of thermodynamics can be adapted to describe black holes\cite{Bekenstein, Hawking1, Hawking2}, there have been repeated attempts to provide a microscopic description of black hole horizons, and of horizons in general.

There have also been attempts to find an \textit{effective} theory of horizon microstates, that can describe the degrees of freedom of the horizon without reference to the underlying theory of quantum gravity. This suggestion is particularly plausible because of the universal appearance of conformal symmetry in the neighborhood of a horizon\cite{Carlip, Dreyer, Silva, Koga, Kang, Medved1, Medved2}, which indicates that the dynamics of a horizon will be governed by a two-dimensional conformal field theory (CFT). Carlip has suggested that the degrees of freedom of this theory are diffeomorphisms that become dynamical at the horizon due to the presence of boundaries or constraints\cite{CarlipLec, Carlip21}. Dynamical actions have been derived for such ``would-be gauge'' degrees of freedom both at spatial infinity in $\AdS$\cite{CarlipAdS1, CarlipAdS2} and $\mathrm{AdS}_5$\cite{Aros}, and at the horizon of the (2+1)-dimensional BTZ black hole\cite{Carlip21}, and these actions are indeed found to describe conformal theories.

In \cite{Mine}, we derived a dynamical action for the near-horizon gravitational degrees of freedom that arise for a very general class of horizons. This result was obtained by imposing physically motivated boundary conditions that preserved the existence and the essential characteristics of a horizon, and identifying the infinitesimal (i.e. first order) diffeomorphisms that preserved these conditions. We then derived a dynamical action for the gravitational degrees of freedom corresponding to these diffeomorphisms from the Einstein-Hilbert action. The resulting action is similar to that of Liouville theory, and the equation of motion derived from this action approaches that of a free two-dimensional conformal field in the near-horizon region.

In this paper we study the thermodynamic properties of the theory found in \cite{Mine}. First we use the fact that the gravitational degrees of freedom at a horizon exhibit a two-dimensional conformal symmetry to calculate the entropy of the horizon. Using the Cardy formula, we find that we can reproduce the Bekenstein-Hawking entropy. This work differs in two main ways from earlier results\cite{Solodukhin, DiasLemos, Giacomini} that find a Liouville action for the gravitational degrees of freedom at a horizon, and use this action to derive the Bekenstein-Hawking entropy. Firstly, our result is valid in an arbitrary number of dimensions, and does not require the horizon to be spherically symmetric. Secondly, instead of obtaining the action by an \textit{ad hoc} dimensional reduction of the Einstein-Hilbert action, or by choosing the action so that it will lead to the expected equation of motion for a Liouville field, our work directly relates the gravitational degrees of freedom at the horizon to the diffeomorphisms that preserve horizon boundary conditions. We only integrate over the spatial coordinates as a final step, after determining that the leading order dynamics are in the $r-t$ plane.

A derivation of the Bekenstein-Hawking entropy is a useful criterion for judging the validity of a theory that describes horizon microstates: however, it is far from being a proof that the theory is correct. Another valuable indicator is seeing whether or not the theory predicts the emission of Hawking radiation. Many different methods have been devised for deriving Hawking radiation\cite{Hawking1, Hawking2, Instantons,Christensen, RobinsonWilczek}. One common feature of all these works is that they analyze quantum matter fields in a classical black hole background, and derive Hawking radiation as a consequence of quantum fields living in a curved space. In order to have a complete picture of horizon thermodynamics, it is important to do the reverse: couple quantized gravitational degrees of freedom to classical matter, and produce the blackbody spectrum of Hawking radiation.

A few steps have been taken in this direction: in \cite{Solodukhin2}, it was shown that Hawking radiation in the near-horizon region could be modeled using a Liouville conformal field theory. Similarly, in \cite{RodriguezYildrim}, an ansatz Liouville theory was proposed to describe the near-horizon region, and this theory was then used to derive the flux of Hawking radiation from several classes of black holes. An alternative approach was taken in \cite{EmparanSachs}, where the Liouville theory of diffeomorphism degrees of freedom at the spatial infinity of $\AdS$ was coupled to scalar field matter. It was shown that the decay rate of the BTZ black hole exactly matched the spectrum of Hawking radiation, including greybody factors. In this work we follow a similar approach, but instead of coupling a classical scalar field to a conformal field theory at the boundary, we identify a coupling in a neighborhood of the horizon, and obtain the Hawking radiation spectrum.

This paper uses the method proposed in \cite{EmparanSachs}, by coupling the horizon degrees of freedom found in \cite{Mine} to scalar field matter and deriving the thermal spectrum of Hawking radiation. Our approach differs from those of \cite{Solodukhin2, RodriguezYildrim} as these earlier papers proposed an ansatz Liouville theory to model horizon phenomena. In contrast, this paper uses a theory described by a dynamical action that was directly derived from the Einstein-Hilbert action by relating the gravitational degrees of freedom at the horizon to the diffeomorphisms preserving the existence and characteristics of the horizon.\cite{Mine}. Our work extends the results of \cite{EmparanSachs}, as our result is valid in an arbitrary number of dimensions, and covers a large class of black holes, not just those black holes whose near-horizon region is $\AdS$. Moreover, the coupling between the gravitational degrees of freedom and the matter fields occurs in the near-horizon region, not at spatial infinity as in \cite{EmparanSachs}.

This paper is structured as follows. In Sec. \ref{sec-CFT} we review the main result found in \cite{Mine}, as well as the formalism necessary to understand the results in this paper. In Sec. \ref{sec-Entropy} we derive the Bekenstein-Hawking entropy of a general class of horizons using the conformal field theory of the horizon degrees of freedom. In Sec. \ref{sec-Hawking} we show how the spectrum of Hawking radiation from a horizon can be reproduced by coupling classical scalar field matter to these gravitational degrees of freedom. We conclude in Sec. \ref{sec-conc}.

\section{Conformal Field Theory at the Horizon}\label{sec-CFT}

We first review some of the notation and concepts used to study the diffeomorphism degrees of freedom at the horizon in \cite{Mine}, and then summarize the main result of that paper.

\subsection{Isolated Horizons}

Our horizon boundary conditions are based on the notion of \textbf{weakly isolated horizons} (WIHs)\cite{Ashtekar, Booth}. Isolated horizons are null sub-manifolds $\Delta$ of spacetime, with an intrinsic metric $q_{ab}$ that is the pull-back of the spacetime metric to $\Delta$. A tensor $q^{ab}$ on $\Delta$ is defined to be an inverse of $q_{ab}$ if it satisfies $q_{am}q_{bn}q^{mn}=q_{ab}$. The inverse is not unique, but all of the definitions and constructions in the isolated horizon formalism are independent of the choice of inverse. Given a null normal $l^\mu$ to $\Delta$, the \textit{expansion} $\theta_{(l)}$ of $l^\mu$ is defined to be
\begin{equation}
\theta_{(l)} := q^{ab}\nabla_a l_b.
\end{equation}
A weakly isolated horizon (WIH) is a sub-manifold $\Delta$ of a spacetime that satisfies the following conditions:
\begin{enumerate}
\item $\Delta$ is topologically $S^2\times\mathbb{R}$ and null,

\item Any null normal $l^\mu$ of $\Delta$ has vanishing expansion, $\theta_{(l)}=0$, and

\item All equations of motion hold at $\Delta$ and the stress energy tensor $T_{\mu\nu}$ is such that $-T^\mu_\nu l^\nu$ is future-causal for any future directed null normal $l^\mu$.
\end{enumerate}
Note that if Condition 2 holds for one null normal to $\Delta$, then it holds for all. WIHs generalize the definitions of Killing horizons and apparent horizons, with the normal vector $l^\mu$ being analogous to a Killing vector, and the requirement that $\exo=0$ clearly being inspired by the notion of trapped surfaces. However, the definition of a WIH is given only \textit{at} the horizon, and does not require a Killing vector to exist even within an infinitesimal neighborhood. Thus, WIHs allow for much greater freedom in the dynamics of the matter and spacetime outside the horizon, while preserving the essential characteristics of the horizon itself.

The surface gravity of an isolated horizon is not uniquely defined. However, given a normal vector $l^\mu$ to $\Delta$, there is a function $\kappa_l$ such that
\begin{equation}\label{eq-sgra}
l^\mu\nabla_\mu l^\nu = \kappa_l l^\nu.
\end{equation}
It is always possible to choose a normal vector $l^\mu$ such that the corresponding $\kappa_l$ is constant everywhere on $\Delta$. Therefore, $\kappa_l$ may be interpreted as the surface gravity of the horizon corresponding to $l^\mu$.

\subsection{Gaussian Null Coordinates and Conformal Coordinates}

In this work we use the system of \textbf{Gaussian null coordinates} (denoted ``GN coordinates'') that are analogous to Eddington-Finkelstein coordinates in Schwarzschild spacetime\cite{Wald2}, and which are well suited for studying horizons as they are adapted to null hypersurfaces. In the neighborhood of any smooth null hypersurface $\Delta$, we can define coordinates $(u,r,x^i)$ such that the metric takes the form:
\begin{equation}\label{eq-GN}
{\rd{s}}^2 = rF{\rd{u}}^2 + 2\rd{u}\rd{r}+2rh_i\rd{u}\rd{x^i}+g_{ij}\rd{x^i}\rd{x^j}.
\end{equation}
$g_{ij}$ is positive definite, and $F, h_i,$ and $g_{ij}$ are smooth functions of $(u,r,x^i)$ that can be expanded in powers of $r$. The null hypersurface is defined by $r=0$, and we have chosen a smooth, non-vanishing vector field $l^\mu$ that is normal to $\Delta$, so that the integral curves of $l^\mu$ are the null geodesic generators of $\Delta$ and we have $l^\mu= (\partial/\partial u)^\mu$ on $\Delta$.

Since an isolated horizon is a null hypersurface, we can construct such a coordinate system in a neighborhood of any isolated horizon. In fact, in the neighborhood of the event horizon of a stationary black hole, or a stationary Killing horizon, we can define these coordinates so that all the metric components are independent of $u$. Extremal Killing horizons correspond to the case where $F$ has a simple root at $r=0$, so that it has the form $F=rf(u,r,x^i)$. We will consider only \textit{non-extremal} horizons, such that $F|_{r=0}\neq 0$. We can then define a non-zero surface gravity for the horizon, using the definition (\ref{eq-sgra}) and taking the normal vector to the $r=0$ hypersurface to be $l^\mu := g^{r\mu}$. The surface gravity $\kappa$ associated with this normal vector is $-\frac{1}{2}F|_{r=0}$, and is related to the \textbf{inverse Hawking temperature} $\beta_H$ of the horizon by $\kappa = \frac{1}{\beta_H}$.

We will also use the coordinate $\tilde{r} := -\ln r$, which puts the metric in the form
\begin{align}\label{eq-Tort}
{\rd{s}}^2 &= e^{-\tilde{r}}\left (F{\rd{u}}^2 - 2\rd{u}\rd{\tilde{r}}+2h_i\rd{u}\rd{x^i}\right ) +g_{ij}\rd{x^i}\rd{x^j},
\end{align}
where $F, h_i,$ and $g_{ij}$ are now taken to be functions of $\tilde{r}$. The horizon is now located at $\tilde{r}\to\infty$. We call these coordinates \textbf{tortoise Gaussian null coordinates}. Finally, we may define conformal coordinates $(\p,\m)$ in terms of $(u,r)$ such that the metric takes the form
\begin{equation}\label{eq-conf2}
{\rd{s}}^2 = 2\g\left (\rd{\p}\rd{\m}+h_{+i}\rd{x^+}\rd{x^i}\right)+g_{ij}\rd{x^i}\rd{x^j},
\end{equation}
and $\g$ has a simple root at $r=0$, so it can be written in the form $\g = e^{\sigma(u,r,x^i)}$, with:
\begin{align}
\sigma(u,r,x^i) = \ln r + \sigma_0(u,x^i) + O(r).
\end{align}
In order to obtain the results in \cite{Mine}, it is necessary to impose one more restriction on $\sigma(u,r,x^i)$: we require that $\partial_i\partial_+\sigma=O(r)$, so that $\sigma = \ln r + \sigma_0(u) + \sigma_1 (x^i) + O(r)$. As the form of the functions $h_{+i}$ in (\ref{eq-conf2}) are left unrestricted, this still allows us to describe a very large class of horizon metrics. For example, all stationary horizons satisfy these conditions.

When defining the conformal coordinates $(\p,\m)$, without loss of generality we can impose
\begin{align}\label{eq-xuRel}
\partial_- u = 0,\quad\partial_r \p = 0.
\end{align}
We can also determine the useful relations:
\begin{align}\label{eq-useRel}
\partial_+r &= O(r),\quad\partial_-r = O(r).
\end{align}

\subsection{An Effective Theory of Gravitational Degrees of Freedom at a Horizon}\label{sec-Action}

In this section we review the derivation given in \cite{Mine} of the dynamical action for the gravitational degrees of freedom at a horizon.

We begin by assuming that we have a weakly isolated horizon $\Delta$ in our spacetime, and that we can define GN coordinates in a neighborhood of $\Delta$ so that the horizon lies at $r=0$ and the metric takes the form (\ref{eq-conf2}). This is the background metric, $\met$, which is taken to be fixed and non-dynamical: the dynamical variables in our setup will be the fluctuations about $\met$ that preserve the existence and basic properties of the horizon $\Delta$. Our aim is firstly to isolate and identify these fluctuations, and secondly to determine their dynamics in the near-horizon region, by deriving their Lagrangian. Note that we do \textit{not} require $\met$ to satisfy Einstein's equations (EEs), but we do require it to satisfy certain constraints that are necessary, but not sufficient, such that it \textit{can} satisfy the EEs as $r\to 0$.

In order to find the relevant excitations about $\met$, we use the definition of WIHs to formulate physically motivated boundary conditions that preserve the existence and characteristics of a horizon. We then identify the most general diffeomorphisms $\xi$ that preserve these boundary conditions, and apply the diffeomorphism to $\met$ to obtain a new metric $\met' := \met + \mathcal{L}_{\xi}\met$. As we wish to study the dynamics of gravitational excitations corresponding to $\xi$, we define the field $\phi$ by $\g' = (1+\phi)\g$. Thus $\phi$ becomes the dynamical field in the problem.

We then evaluate the Einstein-Hilbert action for $\met'$. The transformation $\met' := \met + \mathcal{L}_{\xi}\met$ is an infinitesimal diffeomorphism, i.e. a diffeomorphism to first order in $\xi$. Therefore, as the E-H action is diffeomorphism invariant, at first order the action changes only by a boundary term. However, to higher order, this transformation is not a diffeomorphism: that is, if we set $\met' = \met + h_{\mu\nu}$ for some perturbation $h_{\mu\nu}$ and expand the Einstein-Hilbert action to second order or higher in $h_{\mu\nu}$, then evaluating the E-H action for $\met'$ with $h_{\mu\nu} = \mathcal{L}_{\xi}\met$ will give a non-trivial \textit{bulk} contribution to the action as long as the background metric $\met$ is not required to satisfy the EEs\cite{WilczekPorf}. Thus, in order to obtain a non-trivial action for the gravitational degrees of freedom in the near-horizon region, in this work we consider excitations of $\met$ that are derived from diffeomorphisms (in that they are diffeomorphisms to first order), and compute higher order contributions to the E-H action due to these excitations.

As we are working in the near-horizon region $r\to 0$, and we are considering infinitesimal diffeomorphisms (i.e. small $\phi$), we work to leading order in $(r,\phi)$. Imposing the aforementioned constraints on $\met$ that are necessary, but not sufficient, for it to be able to satisfy the EEs as $r\to 0$, we identify the leading terms in the action as $r\to 0$, which give a non-trivial action for $\phi$. This action is similar to that of Liouville theory, and the equation of motion of the gravitational degrees of freedom approaches that of a free two-dimensional conformal field in the near-horizon region.

We now give the details of the calculation (note that in the remainder of this paper, the term ``diffeomorphism'' always refers to a diffeomorphism to first order.) The first step is to define appropriate boundary conditions that preserve a weakly isolated horizon $\Delta$ at $r=0$ in our spacetime, with a background metric of the form (\ref{eq-conf2}). In order for $\Delta$ to be a WIH, it should have zero expansion $\exo=0$ for all normal vectors $l^\mu$. In GN coordinates, the requirement that the horizon satisfy $\exo=0$ is equivalent to saying that $\partial_u g_{ij} = O(r)$. We then impose conditions that preserve the essential characteristics of the horizon by demanding that, after applying a diffeomorphism:
\begin{enumerate}
\item There is still a null hypersurface at $r=0$. This is equivalent to saying that $g_{uu}'$ (or, in conformal coordinates, $\g'$) has a simple root at $r=0$. \label{cond-1}

\item In conformal coordinates, the metric remains in the form given by Eq.(\ref{eq-conf2}), with $\g' = e^{\sigma'(u,r,x^i)}$ for some $\sigma'(u,r,x^i) = \ln r + \sigma_0'(u) + \sigma_1'(x^i) + O(r)$.\label{cond-2}

\item The induced metric on the $r=0$ hypersurface is preserved, so that $g'_{ij} = g_{ij} + O(r)$.\label{cond-3}. This also ensures that the null hypersurface continues to satisfy $\exo=0$.
\end{enumerate} 

Each of these boundary conditions is physically motivated. Condition \ref{cond-1} is necessary (but not sufficient) for a WIH to still exist at $r=0$ following the diffeomorphism. Condition \ref{cond-2} reflects the fact that the class of horizons we are currently considering can all be written in the form (\ref{eq-conf2}): thus, if the horizon remains intact after the diffeomorphism then the metric should remain in the same form up to trivial diffeomorphisms. This condition is analogous to the boundary conditions defining asymptotically AdS spaces. The metric of any asymptotically AdS space can be written in a special form, called the Fefferman-Graham form\cite{FeffermanGraham}. Therefore the diffeomorphisms that keep the metric in Fefferman-Graham form are the dynamical degrees of freedom at the spatial infinity of an asymptotically AdS space\cite{CarlipAdS1, CarlipAdS2, Aros}. Finally, Condition \ref{cond-3} is derived from the requirement that $\exo=0$ for WIHs. In GN coordinates this requirement is equivalent to $\partial_u g_{ij}=O(r)$. It follows that the induced metric on the $r=0$ hypersurface is a characteristic of the horizon, as it remains constant along $\Delta$. We therefore impose that this characteristic of the horizon is preserved under the diffeomorphism $\xi$, so that $g_{ij}' = g_{ij} + O(r)$. This conditions also ensures that the null hypersurface at $r=0$ continues to satisfy $\exo=0$, so that we still have a WIH at $r=0$ following the diffeomorphism.

In conformal coordinates, the diffeomorphisms preserving these conditions have the form:
\begin{align}\label{eq-diff}
\xi^+ &= \xi^+(\p) + O(r)\\
\xi^- &= \xi^-(\m) + O(r)\nonumber\\
\xi^i &= O(r)\nonumber
\end{align}
We see from the form of $\xi$ that the $(\p,\m)$ coordinates define a natural two-dimensional submanifold where our CFT will live, with infinitesimal conformal transformations being given by $\p \to\p+\xi^+(\p)$, $\m \to\m+\xi^+(\m)$. 

After applying a diffeomorphism of the form (\ref{eq-diff}), the boundary conditions ensure that the metric remains in the form (\ref{eq-conf2}). The non-zero components of the new metric $\met' := \met + \mathcal{L}_{\xi}\met$ are:
\begin{align}\label{eq-newMet}
g'_{+-} &= g_{+-}(1+\partial_+\xi^+ +\partial_-\xi^-+\xi^+\partial_+\sigma+\xi^-\partial_-\sigma)\\
g'_{+i} &= \partial_+\xi^+ g_{+i} + \partial_i\xi^-\g + \xi^-\partial_-g_{+i} + \xi^+\partial_+g_{+i}\nonumber\\
g'_{ij} &= g_{ij} + \xi^+\partial_+g_{ij} +\xi^-\partial_-g_{ij}\nonumber
\end{align}
We now evaluate the Einstein-Hilbert action 
\begin{equation}
I_{EH} = \frac{1}{16\pi G} \int\rd{^nx}  \sqrt{-g}\,\,(R-2\Lambda)
\end{equation}
for the new metric $\met' := \met + \mathcal{L}_{\xi}\met$ after applying the diffeomorphism, and impose a set of constraints that are necessary, but not sufficient, for $\met$ to satisfy the EEs asymptotically as $r\to 0$ (see \ref{app-3} for details of the computations.) This allows us to isolate the dynamics of the gravitational fluctuations about the background metric. 

We define the field $\phi$ by $\g' = (1+\phi)\g$, so that
\begin{equation}\label{eq-phiDef}
\phi = \partial_+\xi^+ + \partial_-\xi^- + \xi^+\partial_+\sigma + \xi^-\partial_-\sigma.
\end{equation}
Although $\sigma$ (and therefore $\phi$) is a function of $x^i$, the fact that $\partial_i\partial_+\sigma=O(r)$ and $\partial_i\partial_-\sigma=O(r)$ means that to leading order $\phi$ is independent of $x^i$ and thus may be considered as a field $\phi(\p,\m)$ on the $(\p,\m)$ submanifold. 

We then evaluate the Einstein-Hilbert action for the new metric $\met'$ to leading order in $(r,\phi)$. We find that all the dynamics in the $x^i$ coordinates disappear to $O(r^2)$, so we may integrate over these coordinates, and the Einstein-Hilbert action takes the form:
\begin{equation}\label{eq-action}
I_{\mathrm{hor}} = \frac{\ar}{16\pi G} \int\rd{^2x} \sqrt{-\hat{g}}\,\, \left (\partial_a\phi\partial^a\phi -\phi\hat{R}+ \lambda (1+\phi)\right )
\end{equation}
where $\ar$ is the cross-sectional area of the horizon, $\hat{g}$ is the induced metric on the $(\p,\m)$ submanifold, and the Ricci scalar $\hat{R}$ is computed from $\hat{g}$. The parameter $\lambda$ is given by
\begin{align}
\lambda := \frac{1}{a_{\Delta}} \int \mathrm{d}^{n-2}x^i\,\,\, \tilde{\lambda}\sqrt{\tilde{g}},
\end{align}
where $\tilde{\lambda} := \frac{4\Lambda e^{\sigma_1(x^i)}}{n-2}$, and $\tilde{g}_{ij} := g_{ij}|_{r=0}$ is the induced metric on the $r=0$ hypersurface.

The equation of motion for the field $\phi$ has the form:
\begin{align}
\partial_+\partial_-\phi + O(r) = 0,
\end{align}
so that $\phi$ becomes a free two-dimensional conformal field in an infinitesimal neighborhood of the horizon.

Note: the above result requires some conditions on the fall-off behavior of the matter energy-momentum (EM) tensor as $r\to 0$ (for details, see \ref{app-3}.) If these conditions are not met, then the potential term for $\phi$ in the action becomes modified, and we need to know the precise form of $T_{\mu\nu}$ in order to find an explicit form for the action. However, the dynamics of the field $\phi$ are unchanged by the addition of matter, as the non-kinetic terms in the action are multiplied by $\sqrt{-\hat{g}}$, which is $O(r)$. Thus these terms become irrelevant as $r\to 0$, so that $\phi$ still becomes a free two-dimensional conformal field in an infinitesimal neighborhood of the horizon. This observation will be important in Sec. \ref{sec-Entropy} and \ref{sec-Hawking}, when we study the thermodynamic properties of the theory described by the action (\ref{eq-action}).

\section{Horizon Entropy}\label{sec-Entropy}

We now calculate the entropy of a large class of horizons using the Cardy formula, a remarkable result that allows us to determine the entropy of any system with a two-dimensional conformal symmetry. The Cardy formula states that, given any unitary two-dimensional conformal field theory with generators $L_n^{\pm}$ of conformal transformations, the asymptotic density of states at eigenvalues $\Delta^{\pm}$ of $L_0^{\pm}$ is given by:
\begin{equation}
\ln \rho(\Delta^+, \Delta^-) \sim 2\pi\sqrt{\frac{c^+\Delta^+}{6}}+2\pi\sqrt{\frac{c^-\Delta^-}{6}}
\end{equation}
This formula has often been used to calculate the horizon entropy of black holes, simply by using a knowledge of the symmetries that govern the gravitational degrees of freedom in a spacetime. Brown and Henneaux\cite{BrownHenneaux} showed that the generators of the asymptotic symmetry group of $AdS_3$ formed a Virasoro algebra with a nontrivial central charge. Strominger then used this result to show that the entropy of $AdS_3$ black holes could be reproduced by using this central charge in the Cardy formula\cite{Strominger}.

This approach has been extended by others \cite{Carlip, Dreyer, Silva, Koga, Kang}, who have imposed physically motivated boundary conditions at a horizon or at spatial infinity to identify the symmetries that preserve these boundary conditions and thus govern the gravitational degrees of freedom in the spacetime. Then, computing the algebra of charges corresponding to these symmetries, and a corresponding central extension (if one exists) has allowed the Bekenstein-Hawking entropy to be reproduced for several types of black holes. The striking feature of this method is that a \textit{classical} conformal symmetry can be enough to determine the entropy, without knowing anything about the underlying quantum theory, lending support to the notion that there is an effective description of the horizon degrees of freedom that is independent of the true theory of quantum gravity. Moreover, the appearance of the Virasoro algebra suggests that this effective theory is a two-dimensional conformal field theory (CFT).

As we have found that the gravitational degrees of freedom at a horizon have an effective description as a 2D CFT, we can also apply the Cardy formula to calculate the entropy of the horizon. As in the works cited above, we have identified the diffeomorphisms $\xi^+$ and $\xi^-$ given in (\ref{eq-diff}) that preserve the existence and characteristics of a horizon, and we have derived a dynamical action for the gravitational fluctuations in the near-horizon region that correspond to these diffeomorphism degrees of freedom. We can therefore compute the algebra of charges corresponding to the symmetries $\xi^+, \xi^-$ of the theory. Our work is close in spirit to that of Solodukhin\cite{Solodukhin} and Giacomini and Pinamonti\cite{Giacomini}.

The energy-momentum tensor derived from $I_{\mathrm{hor}}$ is
\begin{align}
T_{ab} &= \frac{\ar}{16\pi G} \Bigl [ \partial_a\phi\partial_b\phi - \hat{g}_{ab} \left ( \frac{1}{2}\partial_a\phi\partial^a\phi + \frac{\lambda}{2}(1+\phi)\right )\nonumber\\
&\hspace{1.7cm}-(\hat{g}_{ab}\Box\phi - \nabla_a\nabla_b\phi)\Bigr ]
\end{align}
The first step in calculating the horizon entropy is to compute the algebra of charges corresponding to the symmetries of the CFT. The symmetries are given by the diffeomorphisms $\xi^+(x^+)$ and $\xi^-(\m)$. In order to obtain a countable set of charges, we impose a cutoff scale $l$ and define
\begin{equation}
\zeta^\pm_n = \frac{l}{2\pi}e^{\pm\frac{2\pi i}{l} nx_\pm}
\end{equation}
The generators of the corresponding conformal transformations are:
\begin{align}
L_n^{\pm} &= \int_{-l/2}^{l/2} \rd{x_{\pm}}\,\zeta_n^{\pm} T_{\pm\pm}\nonumber\\
&= \frac{\ar}{16\pi G} \frac{l}{2\pi}\int_{-l/2}^{l/2} \rd{x_{\pm}}e^{\pm\frac{2\pi i}{l} nx_\pm} \left [(\partial_{\pm}\phi)^2 +\partial_{\pm}^2\phi\right ]
\end{align}
To evaluate the algebra of charges, we define coordinates $(t,\rho)$ such that $\p = t + \rho$, $\m = t - \rho$, and define $t$ to be the time coordinate. We see from the action in (\ref{eq-action}) that the canonical momentum $\Pi$ conjugate to the field $\phi$ is $\Pi := \delta\mathcal{L}/\delta(\partial_t \phi) = \frac{\ar}{16\pi G} \partial_t\phi$. We thus obtain the Poisson bracket
\begin{equation}
\{\phi(t, \rho), \partial_t\phi(t, \rho')\} = \frac{16\pi G}{\ar}\delta(\rho - \rho') 
\end{equation}
Using the Poisson bracket to evaluate the algebra of charges, we find that
\begin{equation}
\{ L_n^{\pm}, L_m^{\pm}\} = i(n-m)L_{n+m}^{\pm} + \frac{ic^{\pm}}{12}n^3\delta_{n+m,0}
\end{equation}
with central charges
\begin{equation}
c^{\pm} = \frac{3\ar}{4G}
\end{equation}
Now we simply need to determine the eigenvalues of $L_0^{\pm}$ for a given horizon in order to calculate its entropy.

Note that our calculations are completely independent of the form of the potential for $\phi$ in the action. Thus, even if the fall-off conditions on the matter energy-momentum tensor required to obtain the action (\ref{eq-action}) are not met, and the action becomes modified by an additional potential term for $\phi$, the results in this section remain unchanged.

We would now like to evaluate $L_0^{\pm}$ for the classical horizon configuration: this corresponds precisely to the case when there are no fluctuations about the horizon, so that $\phi = 0$, giving $L_0^{\pm}=0$. This is not a serious problem: it is merely due to the fact that the overall normalization of $L_0^{\pm}$ has not been fixed.

In order to determine the normalization, recall the definition of $\phi$ given in (\ref{eq-phiDef}). We see from (\ref{eq-newMet}) that the metric $g_{\mu\nu}'$ after the diffeomorphism has $g_{+-}' = (1+\phi)g_{+-}$. Since we are considering infinitesimal diffeomorphisms, we have $\met' \approx e^{\phi}\met$. Thus $\phi$ can be interpreted as a fluctuation of the conformal factor of the metric on the $(\p,\m)$ submanifold. Recall that we have $\met = e^{\sigma(u,r,x^i)}$ with $\sigma(u,r,x^i) = \ln r + \sigma_0(u)+\sigma_1(x^i)+O(r)$, and our boundary conditions require $\met' = e^{\sigma'(u,r,x^i)}$ with $\sigma'(u,r,x^i) = \ln r + \sigma_0'(u)+\sigma_1'(x^i)+O(r)$. Thus the $\ln r$ term in $\sigma$ is common to all the horizons we are considering, and we can think of it as a fixed, non-dynamical background around which $\phi$ fluctuates. When $\phi = 0$, the conformal factor of the metric on the $(\p,\m)$ submanifold is simply $\ln r$. The ground state value of $L_0^{\pm}$ is therefore given by evaluating $L_0^{\pm}$ for $\phi$ shifted by this background value.

We can calculate $L_0^{\pm}$ for $\phi = \ln r$ by writing $\ln r$ in terms of the coordinates $(\p,\m)$. Consider a general metric of the form (\ref{eq-GN}). As previously stated, such a metric has a well-defined inverse Hawking temperature $\beta_H = \frac{1}{\kappa}$, where $\kappa = -\frac{1}{2}F_{r=0}$. By defining $\p = u$ and $\m = u - \beta_H\ln r$, we can put this metric in the conformal form (\ref{eq-conf2}). It follows that $\ln r = \frac{1}{\beta_H}(\p - \m)$. Evaluating $L_0^{\pm}$ for the solution $\phi = \ln r = \frac{1}{\beta_H}(\p - \m)$ gives:
\begin{equation}
L_0^{\pm} = \frac{\ar l^2 }{8\pi^2 G\beta_H^2}
\end{equation}
We can now calculate the horizon entropy using the Cardy formula, and find:
\begin{equation}
S_H = \frac{\ar l}{4G\beta_H}
\end{equation}
We obtain the desired result of $S_H = \frac{\ar}{4G}$ with $l = \beta_H$, which is a natural choice for $l$ as $\beta_H$ is the period for a thermal ensemble after analytically continuing to imaginary time.

\section{Hawking Radiation}\label{sec-Hawking}

We now investigate another aspect of horizon thermodynamics: Hawking radiation. We couple the gravitational degrees of freedom in the near-horizon region to a classical scalar field, and show that we can produce Hawking radiation from this coupling. Unlike most derivations of Hawking radiation, we will quantize the gravitational theory, and treat the scalar field as a classical background, as in \cite{EmparanSachs}. Our calculations extend and modify those given in \cite{EmparanSachs}, as we work in an arbitrary number of dimensions and couple the scalar field to the gravitational degrees of freedom at the horizon, rather than spatial infinity.

We know from the results described in Sec. \ref{sec-CFT} that the gravitational degrees of freedom in a neighborhood of a horizon can be described by a 2D CFT. We also know how the metric components change under the infinitesimal conformal transformations $x_+ \to x_+'(x_+)$ and $x_- \to x_-'(x_-)$ of the CFT. The fields and operators of the CFT are the components of the metric and objects constructed from the metric, since the metric is the only dynamical field in our framework. Note that this result remains unchanged even if the matter energy-momentum tensor does not satisfy the fall-off conditions that lead to the precise form of the action in (\ref{eq-action}), as $\phi$ still becomes a free two-dimensional conformal field in an infinitesimal neighborhood of the horizon.

There is one subtle point that must be taken into account when quantizing this CFT. When calculating the classical charges $L_0^{\pm}$ in Sec. \ref{sec-Entropy}, we evaluated the charges by shifting $\phi$ by a fixed background configuration corresponding to a horizon. However, when we quantize the field $\phi$, we are quantizing its fluctuations about this background, which have the form given in (\ref{eq-phiDef}). We can see from the relations (\ref{eq-xuRel}) and (\ref{eq-useRel}) that $\partial_-\phi = O(r)$ for these fluctuations. It follows that $L_n^{-} = 0$ for $n \neq 0$, and thus when we quantize the 2D CFT, we end up with only a chiral half of the original theory, with infinitesimal conformal transformations $x_+ \to x_+'(x_+)$.

To determine the conformal weight\cite{DiFrancesco} of the fields and operators in the chiral CFT, we note that as $x_+ = x_+(u)$ and $u = u(x_+)$, infinitesimal transformations of $u$ are equivalent to transformations of $x_+$. Conversely, since $\partial_{r} x_+ = 0$, we find that the coordinate $r$ does not transform under transformations of $x_+$. It follows that the conformal weight of a metric component is equal to the number of lower $u$ indices it has. For example, $g_{ur}$ has conformal weight 1. This result still applies when we switch from using GN coordinates to the tortoise GN coordinates given in (\ref{eq-Tort}).

When we consider adding matter to the system, the total action is the sum of the matter action and the Einstein-Hilbert action. Since the metric appears in the matter action, and $\phi$ is a gravitational degree of freedom, we see that the matter action gives rise to a coupling between $\phi$ and the matter fields. Since we are taking the matter fields to be classical background fields, this means that the matter action will take the form of an operator constructed from $\phi$ that perturbs the theory described by the original action(\ref{eq-action}). Moreover, this perturbation does not disappear in the limit $r\to 0$, which means that it continues to have an effect even in an infinitesimal neighborhood of the horizon. We can determine the conformal weight of this operator, and thus we find that this perturbation results in Hawking radiation.

In order to determine the form of the coupling between a scalar field $\psi$ and the conformal field $\phi$ that arises in the matter action, we consider the scalar field as a classical background, and quantize the CFT. This means that we take the scalar field on-shell in the bulk of the spacetime. Thus the scalar field action will result in a perturbation of the CFT by adding an operator to the CFT action. In tortoise GN coordinates, the scalar field action in the near-horizon region is
\begin{align}
I_s &= \int_{\mathcal{M}} \sqrt{-g}g^{\mu\nu}\partial_\mu\psi\partial_\nu\psi + \int_{\Delta} \sqrt{-g}g^{\tilde{r}\mu}\psi\partial_\mu\psi
\end{align}
The boundary term is evaluated at the horizon. Inspecting the action, we find that all the terms either go to zero in the near-horizon region as $\tilde{r}\to\infty$, or exhibit a coupling of the classical scalar field to an operator of conformal weight 1. These couplings are given by:
\begin{align}
I_{\mathrm{int}} &= \int_{\mathcal{M}} \sqrt{-g}\left ( g^{\tilde{r}\tilde{r}}\partial_{\tilde{r}}\psi\partial_{\tilde{r}}\psi\right ) + \int_{\Delta} \sqrt{-g}\left ( g^{\tilde{r}\tilde{r}}\psi\partial_{\tilde{r}}\psi\right )\nonumber\\
&\sim \int_{\mathcal{M}} g_{u\tilde{r}}\left ( g^{\tilde{r}\tilde{r}}\partial_{\tilde{r}}\psi\partial_{\tilde{r}}\psi\right )+ \int_{\Delta} g_{u\tilde{r}}\left ( g^{\tilde{r}\tilde{r}}\psi\partial_{\tilde{r}}\psi\right )\label{eq-coupling}
\end{align}
When we quantize the near-horizon CFT, these couplings add a perturbation to the CFT action in the form of an operator $\mathcal{O}(u)$ of conformal weight 1. This perturbation will induce transitions between closely spaced states of the CFT, resulting in the emission of radiation. From the form of the metric in (\ref{eq-Tort}) we see that this perturbation remains finite even in an infinitesimal neighborhood of the horizon, so that our calculation remains valid as $r\to 0$.

We assume that the form of the scalar field is not affected by small deformations of the metric. It is not necessary to know the specific form of $\psi$, but it is instructive to work out a simple example. If the background metric is a spherically symmetric, static metric with a Killing horizon that has the form
\begin{equation}
{\rd{s}}^2 = -f(x){\rd{t}}^2+\frac{1}{f(x)}{\rd{x}}^2 + x^2{\rd{\Omega}}^2,
\end{equation}
with $f(x) = \frac{2}{\beta_H}(x-x_h) + O((x-x_h)^2)$, then we can write this metric in GN coordinates by defining $r:= x-x_h$ and $u:=t+\frac{\beta_H}{2}\ln r$. We can describe $\psi$ as an infinite collection of 2-dimensional scalar fields $\psi_{l,m}$ of the form
\begin{equation}
\psi_{l,m} = e^{i(t-\tilde{r})} + e^{i(t+\tilde{r})}
\end{equation}
where $\tilde{r}:=\frac{\beta_H}{2}\ln r$ is the radial tortoise coordinate\cite{RobinsonWilczek}. In this simple case of a static metric with a Killing horizon, substituting the modes $\psi_{l,m}$ into (\ref{eq-coupling}) gives a coupling that remains finite even infinitesimally close to the horizon.

This coupling will lead to transitions between the states of the CFT, so that the horizon produces Hawking radiation. We can compute the macroscopic decay rate using standard conformal field theory methods, following the approaches of \cite{EmparanSachs} and \cite{MaldacenaStrominger}. The operator $\mathcal{O}(u)$ introduced by the coupling to the scalar field will lead to a transition amplitude between initial and final states of the horizon in the presence of an external flux with frequency $\omega$ of the form:
\begin{equation}
\mathcal{M} \sim \int \rd{u} \langle f | \mathcal{O}(u) | i \rangle e^{-i\omega u}.
\end{equation}
Squaring and summing over final states, we get:
\begin{equation}\label{eq-M}
\sum_f |\mathcal{M}|^2 \sim \int \rd{u}\rd{u'} \langle i | \mathcal{O}(u)\mathcal{O}(u') | i \rangle e^{-i\omega (u-u')}
\end{equation}
for the decay rate. As we have already stated, we can define the surface gravity $\kappa$ (and therefore, the temperature) of any horizon that satisfies our boundary conditions as $\kappa := -\frac{1}{2}F|_{r=0}$. Thus, the class of horizons we are studying are in thermal equilibrium and can therefore be considered as thermal states with a well-defined temperature. We therefore average over the intial states assuming that the distribution is given by a Boltzmann spectrum. If the temperature of the horizon is $T_H$, then the decay rate is given by finite temperature two-point functions, which have the form
\begin{equation}
\langle \mathcal{O}^\dagger(0)\mathcal{O}(u)\rangle_{T_H} \sim \left [\frac{\pi T_H}{\sinh(\pi T_H u)} \right ]^2
\end{equation}
In order to evaluate the integrals in (\ref{eq-M}), we use standard techniques of contour integration and assume that we are calculating emission rates. We find that the emission rate is given by
\begin{equation}
\Gamma \sim \frac{\pi\omega}{e^{\frac{\omega}{T_H}}-1},
\end{equation}
where we have divided by a factor of $\omega$ to account for the normalization of the outgoing scalar. Thus we obtain the familiar blackbody spectrum of Hawking radiation.

\section{Conclusion}\label{sec-conc}

We have investigated the thermodynamic properties of horizons by using the dynamical description of the diffeomorphism degrees of freedom obtained in \cite{Mine}. Using the Cardy formula, we computed the entropy of the horizon and reproduced the expected Bekenstein-Hawking entropy. This result suggests that the classical conformal symmetry imposed by boundary conditions at a horizon is enough to determine the entropy of the horizon, without reference to the underlying theory of quantum gravity. The wide applicability of the result to many kinds of horizons, including cosmological and acceleration horizons\cite{Unruh, GibbonsHawking}, indicates that the universality of the Bekenstein-Hawking entropy formula is results from the fact that a two-dimensional conformal symmetry is always induced near a horizon. We have also provided evidence for the validity of the effective description of horizon degrees of freedom as a 2D CFT by coupling the effective theory to a classical scalar background and showing that this produces Hawking radiation. Our result shows that although the effective theory is not the ``true'' theory of quantum gravity, it can provide a way of quantizing the gravitational degrees of freedom at a horizon.

\begin{appendix}

\section{Computing the Dynamical Action}\label{app-3}

In this section we present the details of the calculations used to derive the dynamical action in Sec. \ref{sec-Action}.

Evaluating the Ricci tensor for metrics of the form (\ref{eq-conf2}), we find:
\begin{align}
R_{+-} &= -\partial_+\partial_-\sigma + O(r)\\
R_{-i} &= \frac{1}{2}(\partial_-^2\sigma h_{+i} + \partial_-^2h_{+i} + \partial_-\sigma\partial_-h_{+i}) + O(r)\\
R_{+i} &= \frac{1}{2}(\partial_-\sigma \partial_+h_{+i} + \partial_+\partial_-h_{+i}) + O(r)
\end{align}
We will assume that the matter energy-momentum tensor in the near-horizon region satisfies:
\begin{align}
T_{+-}  \,\,\,\hat{=}\,\,\, O(r^2)\label{eq-TAssumpGen1}\\
T_{-i}  \,\,\,\hat{=}\,\,\, O(r)\label{eq-TAssumpGen2}\\
T_{+i}  \,\,\,\hat{=}\,\,\, O(r)\label{eq-TAssumpGen3}\\
T_{ij}  \,\,\,\hat{=}\,\,\, O(r)\label{eq-TAssumpGen4}
\end{align}
where ``$\hat{=}$'' indicates that the constraint holds as $r\to0$ (The reasonableness of these constraints is discussed at greater length in \cite{Mine}. If they are not met, then the effective action for the gravitational degrees of freedom in the near-horizon region become modified by a potential term. However, the dynamics described by the action remain unchanged, even in this case.) 

Looking at the Einstein equation:
\begin{equation}\label{eq-EE}
R_{\mu\nu} - \frac{1}{2}Rg_{\mu\nu} + \Lambda g_{\mu\nu} = 8\pi G T_{\mu\nu},
\end{equation}
this gives the following constraints on the metric:
\begin{align}
\partial_+\partial_-\sigma \,\,\,\hat{=}\,\,\, O(r)\label{eq-GenConst1}\\
\partial_-^2\sigma h_{+i} + \partial_-^2 h_{+i} + \partial_-\sigma\partial_- h_{+i} \,\,\,\hat{=}\,\,\, O(r)\label{eq-GenConst2}\\
\partial_-\sigma \partial_+ h_{+i} + \partial_+\partial_- h_{+i} \,\,\,\hat{=}\,\,\, O(r)\label{eq-GenConst3}\\
R - \frac{2n\Lambda}{n-2} \,\,\,\hat{=}\,\,\, O(r)\label{eq-GenConst4}
\end{align}
These conditions are necessary (but not sufficient) for $\met$ to satisfy the Einstein equations as $r\to 0$.

We apply a diffeomorphism $\xi$ of the form (\ref{eq-diff}) to the background metric $\met$, obtaining a new metric $\met' := \met + \diff\met$. We then evaluate the Einstein-Hilbert action for the new metric $\met'$ with the constraints (\ref{eq-GenConst1})-(\ref{eq-GenConst4}) applied to $\met$ in the near-horizon region, thus isolating the gravitational fluctuations about this background that preserve the horizon. When evaluating the Einstein-Hilbert action, everything is calculated from the new metric, including the inverse metric and the metric determinant, as the metric itself is the only dynamical field in the problem. The final form of the action determines the dynamics of the horizon degrees of freedom in an infinitesimal neighborhood of the horizon. As we do not require $\met$ to satisfy the Einstein equations, we get a non-trivial form for the action.

We begin with the Einstein-Hilbert action
\begin{equation}
I_{EH} = \frac{1}{16\pi G} \int\rd{^nx}  \sqrt{-g}\,\,(R-2\Lambda),
\end{equation}
and evaluate the action for the new metric $\met'$. We see from (\ref{eq-newMet}) that this metric has the form $\g' = (1+\phi)\g$ and $g_{ij}' = g_{ij} + O(r)$, with $\phi$ given by
\begin{equation}
\phi = \partial_+\xi^+ +\partial_-\xi^-+\xi^+\partial_+\sigma+\xi^-\partial_-\sigma.
\end{equation}
Although $\sigma$ (and therefore $\phi$) is now a function of $x^i$, the requirement that $\partial_i\partial_+\sigma=O(r)$ and $\partial_i\partial_-\sigma=O(r)$ means that to leading order $\phi$ is independent of $x^i$ and may be considered as a field on the $(\p,\m)$ submanifold. This $\phi$ will end up being the dynamical degree of freedom in the near-horizon region. As we are interested in the near-horizon region $r\to 0$, and we are considering infinitesimal diffeomorphisms, we work to leading order in $(r,\phi)$. To second order in $\phi$, the inverse metric ${g'}^{\mu\nu}$ has the form
\begin{align}
{g'}^{+-} &= (1-\phi + \phi^2)g^{+-}\\
{g'}^{ij} &= g^{ij} + O(r)
\end{align}
We can now begin evaluating each of the terms necessary to compute the Einstein-Hilbert action. First we find that
\begin{align}\label{eq-AppMetDet}
\sqrt{-g'} &= \g'\sqrt{\tilde{g}} + O(r^2)
\end{align}
In order to evaluate the Einstein-Hilbert action for $\met'$, we need to compute
\begin{align}
R' &= 2(g^{'+-}R'_{+-} + g^{'i-}R'_{i-}) + g'^{ij}R'_{ij}
\end{align}
to leading order in $(r,\phi)$. We first find
\begin{align}
R_{+-} &= -\partial_-\Gamma_{++}^+\nonumber\\
&\qquad+ \frac{1}{2} \biggl [g^{ij}(\partial_-\partial_j g_{+i} - \partial_i\partial_j\g - \partial_-\partial_+g_{ij})\nonumber\\
&\qquad\qquad\qquad+ (\partial_-g_{+j} - \partial_j\g)g^{ij}(\tilde{\Gamma}^k_{ki} + \frac{1}{2}g^{k-}\partial_-g_{ik})\\
&\qquad\qquad\qquad+ (\partial_-g_{+j} - \partial_j\g)(\partial_-g^{j-} + \partial_i g^{ij})\nonumber\\
&\qquad\qquad\qquad - g^{i-}(\partial_i\partial_-\g - \partial_-^2 g_{+i}) \biggr ] + O(r^2)\\
R_{-i} &= \frac{1}{2}g^{+-}(\partial_-^2 g_{+i} - \partial_-\partial_i \g) + \frac{1}{2}(g^{+-})^2(\partial_-\g)(\partial_i\g - \partial_- g_{+i})\nonumber\\
&\hspace{0.7cm} + O(r)
\end{align}
where $\tilde{A}$ denotes a quantity $A$ computed with respect to the induced metric $\tilde{g}_{ij} := g_{ij}|_{r=0}$ on the $r=0$ hypersurface. This gives:
\begin{align}
2(g^{i-}R_{i-} + g^{+-}R_{+-})&= g^{i-}g^{+-}(\partial_-^2g_{+i} - \partial_i\partial_-\g)\nonumber\\
&\qquad+ g^{i-}(g^{+-})^2(\partial_-\g\partial_i\g - \partial_-\g\partial_i g_{+i})\nonumber\\
&\qquad -g^{+-}2\partial_-\Gamma_{++}^+\nonumber\\
&\qquad + g^{ij}g^{+-}(\partial_-\partial_j g_{+i} - \partial_i \partial_j \g - \partial_- \partial_+ g_{ij})\nonumber\\
&\qquad + g^{+-}(\partial_- g_{+j} - \partial_-\g)g^{ij}(\tilde{\Gamma}^k_{ki} + \frac{1}{2}g^{k-}\partial_- g_{ik})\nonumber\\
&\qquad + g^{+-}(\partial_-g_{+j} - \partial_-\g)(\partial_i g^{ij} + \partial_- g^{j-})\nonumber\\
&\qquad + O(r)\label{eq-Gen1}\end{align}
Similarly, we can compute:
\begin{align}
g^{ij}R_{ij} &= -g^{ij}g^{+-}(\partial_+\partial_-g_{ij} + \partial_i\partial_j\g)\nonumber\\
&\qquad + \frac{1}{2}g^{ij}g^{+-}(\partial_-\partial_ig_{+j} +\partial_-\partial_j g_{+i})\nonumber\\
&\qquad + \frac{1}{2}g^{ij}(\partial_j\g\partial_i\g - \partial_-g_{+j}\partial_-g_{+i})\nonumber\\
&\qquad + g^{ij}g^{+-}\partial_k\g\tilde{\Gamma}_{ij}^k + g^{ij}\tilde{R}_{ij} + O(r)\label{eq-Gen2}
\end{align}
We can now compute $\g'R'$ by varying all the quantities in (\ref{eq-Gen1}) and (\ref{eq-Gen2}) under a diffeomorphism $\xi$ of the form (\ref{eq-diff}). We write:
\begin{align}
\g'R' &= -2\g'g^{'+-}\partial_-\Gamma_{++}^{'+} + B(\phi, \met)
\end{align}
for some function $B(\phi, \met)$. If we can show that
\begin{align}\label{eq-Key}
B(\phi, \met) - \g'(R + 2g^{+-}\partial_+\partial_-\sigma) = O(r^2),
\end{align}
then we can write:
\begin{align}
\g'R' &= -2\g'g^{+-}\partial_-\Gamma_{++}^{'+} + \g' (R + 2g^{+-}\partial_+\partial_-\sigma) + O(r^2)\nonumber\\
&= 2\g'(-g^{+-}\partial_-\Gamma_{++}^{'+} + g^{+-}\partial_+\partial_-\sigma)\nonumber\\
&\qquad + \g'\frac{2n\Lambda}{n-2} + O(r^2)\label{eq-GenFinalR}
\end{align}
Calculating the first term in the above expression, we find:
\begin{align}\label{eq-AppAction1}
&2\g'(-g^{+-}\partial_-\Gamma_{++}^{'+} + g^{+-}\partial_+\partial_-\sigma)\nonumber\\
&\qquad= 2\g'g^{+-}(-{g'}^{+-}\partial_+\partial_-\g' + ({g'}^{+-})^2\partial_-\g'\partial_+\g')\nonumber\\ 
&\qquad\qquad + 2\g'g^{+-}\partial_+\partial_-\sigma
\end{align}
To evaluate the first two terms in (\ref{eq-AppAction1}) in terms of $\phi$ and the background metric $\met$, we compute:
\begin{align}
-{g'}^{+-}\partial_+\partial_-\g' + ({g'}^{+-})^2\partial_-\g'\partial_+\g' &= \partial_+\phi\partial_-\phi\nonumber\\
&\qquad+ (1+\partial_-\xi^- + \partial_+\xi^+)\partial_+\partial_-\sigma\nonumber\\
&\qquad+ \xi^+\partial_+^2\partial_-\sigma + \xi^-\partial_+\partial_-^2\sigma\nonumber\\
&\qquad+ O(\phi^3, r\phi^2)\label{eq-RicciNewApp}
\end{align}
We are ignoring terms of $O(\phi^3)$ and $O(r\phi^2)$ as we are working to leading order in $(r,\phi)$ and the leading terms in the action will be at most $O(\phi^2)$ or $O(r\phi)$.

We can now write (\ref{eq-AppAction1}) as:
\begin{align}
&2{g'}^{+-}\partial_+\phi\partial_-\phi\label{eq-Kin}\\
&\qquad-2{g'}^{+-}\partial_+\partial_-\sigma(1+\partial_+\xi^+ + \partial_-\xi^-)\label{eq-R1}\\
&\qquad-2{g'}^{+-}(\xi^+\partial_+^2\partial_-\sigma  + \xi^-\partial_+\partial_-^2\sigma)\label{eq-R2}\\
&\qquad+ 2g^{+-}\partial_+\partial_-\sigma + O(r)\label{eq-R3}
\end{align}
Multiplying the first term (\ref{eq-Kin}) of (\ref{eq-AppAction1}) with $\g'$ then gives:
\begin{align}
\sqrt{-\hat{g}} \left (2g^{+-}\partial_+\phi\partial_-\phi\right ) &= \sqrt{-\hat{g}} \left (g^{ab}\partial_a\phi\partial_b\phi\right )\nonumber\\
&= \sqrt{-\hat{g}} \left (\partial_a\phi\partial^a\phi\right )\label{eq-Final1}
\end{align}
where the index $a \in (\p,\m)$ and $\hat{g}_{ab}$ is the induced metric on the $(\p,\m)$ submanifold. 

Similarly, multiplying the terms (\ref{eq-R1})-(\ref{eq-R3}) with $\g'$ gives:
\begin{align}
&-2\left (\xi^+\partial_+^2\partial_-\sigma  + \xi^-\partial_+\partial_-^2\sigma \right )-2\partial_+\partial_-\sigma\left ( 1+\partial_+\xi^+ + \partial_-\xi^-\right ) \nonumber\\
&\qquad+ 2\partial_+\partial_-\sigma(1+\phi)
\end{align}
This expression simplifies to
\begin{align}
&2\left [\partial_+\partial_-\sigma (\xi^+\partial_+\sigma + \xi^-\partial_-\sigma) -\xi^+\partial_+^2\partial_-\sigma  - \xi^-\partial_+\partial_-^2\sigma\right ]\label{eq-Final2a}
\end{align}
We now use integration by parts to rewrite (\ref{eq-Final2a}) as
\begin{align}
&= 2\partial_+\partial_-\sigma (\xi^+\partial_+\sigma + \xi^-\partial_-\sigma + \partial_+\xi^+ + \partial_-\xi^-)\nonumber\\
&= -\left (\g \phi\hat{R}\right)\nonumber\\
&= -\sqrt{\hat{g}}\phi\hat{R}\label{eq-Final2b},
\end{align}
where $\hat{R}$ is the Ricci scalar corresponding to the induced metric $\hat{g}_{ab}$ on the $(\p,\m)$ submanifold, given by
\begin{equation}
\hat{R} = -2g^{+-}\partial_+\partial_-\sigma\label{eq-hatR}
\end{equation}

Thus we can see that (\ref{eq-AppAction1}) is equivalent to
\begin{align}\label{eq-GenFinal1}
\sqrt{-\hat{g}}(\partial_a\phi\partial^a \phi - \phi\hat{R})
\end{align}
Looking at the form of the metric in (\ref{eq-conf2}), we might be concerned that $x^i$-dependence will enter through $\hat{g}$ and $\hat{R}$, when we are trying to define an action on the $(x_+, x_-)$ submanifold. However, recall that $g_{+-} = e^{\sigma(u,r,x^i)}$ with $\sigma = \ln r + \sigma_0(u) + \sigma_1(x^i) + O(r)$. Thus the $x^i$-dependence in $\sqrt{-\hat{g}} = g_{+-}$ cancels with the $x^i$ dependence in the kinetic term $\partial_a\phi\partial^a\phi = 2g^{+-}\partial_+\phi\partial_-\phi$. We have already established that $\phi$ can be interpreted as a field on the $(x_+, x_-)$ submanifold to leading order, so the kinetic term can be defined on the $(x_+, x_-)$ submanifold. A similar cancellation of the $x^i$-dependence occurs for the $\hat{R}$ term, as can be seen from the definition (\ref{eq-hatR}) of $\hat{R}$. As a result, we can interpret the expression (\ref{eq-GenFinal1}) as a quantity defined on the $(x_+, x_-)$ submanifold, by redefining:
\begin{align}
\hat{g}_{+-} := e^{\ln r + \sigma_0(u) + O(r)}
\end{align}
and
\begin{align}
\hat{R} = -2\hat{g}^{+-}\partial_+\partial_-\sigma
\end{align}
Substituting (\ref{eq-GenFinal1}) back into (\ref{eq-GenFinalR}) we find:
\begin{align}
\g'(R' - 2\Lambda) &= \sqrt{-\hat{g}}(\partial_a\phi\partial^a \phi - \phi\hat{R} + \tilde{\lambda}(1+\phi))
\end{align}
where the $x^i$-dependence of $\sqrt{-\hat{g}}$ only appears in the last term, and has been incorporated into $\tilde{\lambda} := \frac{4\Lambda e^{\sigma_1(x^i)}}{n-2}$. This gives:
\begin{align}
\sqrt{-g'}(R' - 2\Lambda) &= \sqrt{\tilde{g}}\sqrt{-\hat{g}}(\partial_a\phi\partial^a \phi - \phi\hat{R} + \tilde{\lambda}(1+\phi))
\end{align}
to leading order in $(r,\phi)$. We can integrate over the $x^i$ coordinates in the Einstein-Hilbert action to obtain the final dynamical action:
\begin{align}
I_{\mathrm{hor}} = \frac{\ar}{16\pi G} \int\rd{^2x} \sqrt{-\hat{g}}\,\, \left (\partial_a\phi\partial^a\phi -\phi\hat{R}+ \lambda (1+\phi)\right )\nonumber
\end{align}
where $\ar$ is the cross-sectional area of the horizon, the variables of integration are $(\p,\m)$, and the parameter $\lambda$ is given by:
\begin{align}
\lambda := \frac{1}{a_{\Delta}} \int \mathrm{d}^{n-2}x^i\,\,\, \tilde{\lambda}\sqrt{\tilde{g}},
\end{align}
where the integral is carried out over the $(n-2)$ coordinates $x^i$.

Now all we have to do is derive (\ref{eq-Key}) in order to obtain our final result. By direct computation of $\g'R'$, we find:
\begin{align}
&B(\phi, \met) - \g'(R + 2g^{+-}\partial_+\partial_-\sigma)\nonumber\\
&\quad= -\phi g^{ij}\tilde{\Gamma}^k_{ij}\partial_k\g + g^{ij}\tilde{\Gamma}_{ij}^k\partial_k(\delta\g)\label{eq-Term3}\\
&\quad-\phi\partial_i g^{ij}(\partial_- (\delta g_{+j}) - \partial_j(\delta \g))+ \partial_i g^{ij}(\partial_i (\delta g_{+j}) - \partial_j(\delta \g))\label{eq-Term9b}\\
&\quad-\phi g^{ij}\tilde{\Gamma}^k_{ki}(\partial_- g_{+j} - \partial_j\g)+ g^{ij}\tilde{\Gamma}^k_{ki}(\partial_i (\delta g_{+j}) - \partial_j(\delta \g))\label{eq-Term4b}\\
&\quad-2\phi g^{ij}(\partial_-\partial_j g_{+i} - \partial_i\partial_j\g - \partial_-\partial_+ g_{ij})\label{eq-Term2a}\\
&\quad+ 2g^{ij}(\partial_-\partial_j (\delta g_{+i}) - \partial_i\partial_j(\delta \g) - \partial_-\partial_+ (\delta g_{ij}))\label{eq-Term2b}\\
&\quad+ \frac{1}{2}g^{ij}g^{+-}(1-\phi)(\partial_j \g'\partial_i\g' - \partial_- g_{+j}'\partial_- g_{+i}')\label{eq-Term5}\\
&\quad- \frac{1}{2}g^{ij}g^{+-}(1+\phi)(\partial_j \g\partial_i\g - \partial_- g_{+j}\partial_- g_{+i})\label{eq-Term5b}\\
&\quad+ 2g^{'i-}(\partial_-^2 g_{+i}' - \partial_-\partial_i \g')- 2g^{i-}(1+\phi)(\partial_-^2 g_{+i} - \partial_-\partial_i \g)\label{eq-Term6b}\\
&\quad+ g^{'i-}g^{+-}(1-\phi)\partial_-\g'(\partial_i\g' - \partial_-g_{+i}')\label{eq-Term7}\\
&\quad- g^{i-}g^{+-}(1+\phi)\partial_-\g(\partial_i\g - \partial_-g_{+i})\label{eq-Term7b}\\
&\quad+ \partial_-g^{'j-}(\partial_-g_{+j}' - \partial_j\g')- \partial_-g^{j-}(1+\phi)(\partial_-g_{+j} - \partial_j\g)\label{eq-Term8b}
\end{align}
We simplify the above expression by applying the constraints (\ref{eq-GenConst1})-(\ref{eq-GenConst3}) to the background metric $\met$. To simplify the terms (\ref{eq-Term3}):
\begin{align}
-\phi\partial_k\g + \partial_k(\delta\g) &= -\phi \partial_k\g + \partial_k(\phi\g)\nonumber\\
&= \g\partial_k\phi = O(r^2),\nonumber
\end{align}
as $\g = O(r)$ and $\partial_k\phi = O(r)$ due to the conditions $\partial_{\pm}\partial_i\sigma = O(r)$. It follows that the terms (\ref{eq-Term3}) combine to give a quantity that is $O(r^2)$.

In order to simplify the terms (\ref{eq-Term9b}) as well as (\ref{eq-Term4b}), consider:
\begin{align}
&-\phi(\partial_- (\delta g_{+j}) - \partial_j(\delta \g))+ (\partial_i (\delta g_{+j}) - \partial_j(\delta \g))\nonumber\\
&\qquad\qquad= \xi^+\g(\partial_-\sigma\partial_+h_{+i} + \partial_-\partial_+ h_{+i})\nonumber\\
&\qquad\qquad\qquad+ \xi^-\g(\partial_-^2\sigma h_{+i} + \partial_-\sigma\partial_- h_{+i} + \partial_-^2 h_{+i})\nonumber\\
&\qquad\qquad= O(r^2)
\end{align}
by the constraints (\ref{eq-GenConst2})-(\ref{eq-GenConst3}) on $\met$. It follows that the terms (\ref{eq-Term9b})-(\ref{eq-Term4b}) combine to give a quantity that is $O(r^2)$. The terms (\ref{eq-Term2a})-(\ref{eq-Term2b}) simplify in the same way to give a quantity that is $O(r^2)$.

We are left with the terms (\ref{eq-Term5})-(\ref{eq-Term8b}). In order to simplify these terms, we use the fact that
\begin{align}
{g'}^{-i} &= -g^{+-}(1-\phi)g'_{+j}g^{ij} + O(r)\\
&= (1-\phi)g^{-i} - (1-\phi)g^{+-}g^{ij}\delta g_{+j} + O(r)
\end{align}
Direct computation and the application of the constraints (\ref{eq-GenConst2})-(\ref{eq-GenConst3}) shows that the terms (\ref{eq-Term5})-(\ref{eq-Term8b}) also combine to give a quantity that is $O(r^2)$. So finally, we find that (\ref{eq-Key}) holds. Note that constraint (\ref{eq-GenConst4}) was not required for these computations.

\end{appendix}

\end{document}